\documentstyle[aps
,epsfig%
,preprint%
]{revtex}
\newcommand{\be}{\begin{equation}}
\newcommand{\ee}{\end{equation}}
\newcommand{\bea}{\begin{eqnarray}}
\newcommand{\eea}{\end{eqnarray}}

\begin{document}
\draft
\preprint{Guchi-TP-003}
\date{\today
}
\title{
Exact Solutions for Boson-Fermion Stars in (2+1) dimensions 
}
\author{Kenji~Sakamoto${}^1$ and
Kiyoshi~Shiraishi${}^{1,2}$%
\thanks{e-mail: {\tt g00345@simail.ne.jp,
shiraish@sci.yamaguchi-u.ac.jp}
}}
\address{${}^1$Graduate School of Science and Engineering, 
Yamaguchi University\\ 
Yoshida, Yamaguchi-shi, Yamaguchi 753-8512, Japan
}
\address{${}^2$Faculty of Science, Yamaguchi University\\
Yoshida, Yamaguchi-shi, Yamaguchi 753-8512, Japan
}
\maketitle
\begin{abstract}
We solve Einstein equations coupled to a complex scalar field with
infinitely large self-interaction, degenerate fermions, and a negative
cosmological constant in $(2+1)$ dimensions.  
Exact solutions for static
boson-fermion stars are found when circular symmetry is assumed.
We find that the minimum binding energy of boson-fermion star
takes a negative value if the value of the cosmological constant is
sufficiently small.
\end{abstract}


\section{introduction}

Lower-dimensional gravity provides us with a useful arena where
we can often construct exactly solvable models and can study the
model analytically.
The solutions for relativistic boson stars (\cite{BS}\cite{LM} for
reviews) are only numerically obtained in four dimensions.
In the previous paper~\cite{KS}, we obtained an exact solution for a
nonrotating boson star in
$(2+1)$ dimensional gravity with a negative cosmological constant. We
assumed that the scalar field has a strong  self-interaction.
An infinitely large self-interaction term in the model leads to much
simplication as in the $(3+1)$ dimensional case~\cite{CSW}.


In the present paper, we derive an exact solution for a nonrotating 
boson-fermion star in $(2+1)$ dimensional gravity with a negative
cosmological constant. We assume that the scalar field has a strong 
self-interaction as in the previous paper.

Four-dimensional boson-fermion stars have been studied by many 
authors~\cite{BFS}.
Our three-dimensional model includes an additional constant,
the cosmological constant.
We will see that the value of the cosmological constant is
restricted by the stability condition of the boson-fermion star.
We will also find, nevertheless, that some physical quantities
of the boson-fermion star behave as in the four-dimensional case.

In Section~\ref{sec:2}, we show the field equations which we solve.
In Section~\ref{sec:3}, we obtain exact solutions for the field
equations shown in Section~\ref{sec:2}. We pay our attention on
the expression of mass of the boson-fermion star.
The particle numbers of bosons and fermions for the boson-fermion star
described by the exact solutions is given in Section~\ref{sec:4}.
In Section~\ref{sec:5}, we restrict our attention to the fermion star,
i.e., the case that the particle number of bosons vanishes.
We discuss stablity of the  boson-fermion star in Section~\ref{sec:6}.
Section~\ref{sec:7} is devoted to conclusion.

\section{field equations}
\label{sec:2}

We consider a complex scalar field $\varphi$ with mass $m_B$ and a
quartic self-coupling constant $\lambda$. We assume that the coupling
constant takes an infinitely large value.
The equation of motion for the scalar field is
\begin{equation}
\nabla^2\varphi-m_B^2\varphi-\lambda\left|\varphi\right|^2\varphi=0.
\label{eq:KG}
\end{equation}

In addition, we assume that a degenerate fermion field with mass $m_F$
is coupled to gravity. In this paper, we assume no direct coupling
between the boson and fermion fields.


The Einstein equation is written as
\begin{equation}
R^{\mu}_{\nu}-\frac{1}{2}\delta^{\mu}_{\nu}R=
8\pi G\left(T^B{}^{\mu}_{\nu}+T^F{}^{\mu}_{\nu}\right)+
C\delta^{\mu}_{\nu},
\label{eq:EE}
\end{equation}
where the positive constant
$C$ stands for the (negative) cosmological constant.
$G$ is the Newton constant.


The energy-momentum tensor of the boson field is
\begin{equation}
T^B{}^{\mu}_{\nu}=
2{\rm Re}\left(\nabla^{\mu}\varphi^{*}\nabla_{\nu}\varphi-
\frac{1}{2}\left|\nabla_{}\varphi\right|^2
\delta^{\mu}_{\nu}\right)-
m_B^2\left|\varphi\right|^2\delta^{\mu}_{\nu}-\frac{\lambda}{2}
\left|\varphi\right|^4\delta^{\mu}_{\nu}.
\label{eq:TB}
\end{equation}


The energy-momentum tensor of the fermion field is written
in the perfect fluid form as
\begin{equation}
T^F{}^{\mu}_{\nu}=diag.\left(-\rho,P,P\right),
\label{eq:TF}
\end{equation}
where the hydrostatic equilibrium is assumed.

In three dimensions, the energy density and pressure
are given by~\cite{FTD}
\begin{equation}
\rho=\frac{m_F^3}{6\pi}\left(\frac{\mu^3}{m_F^3}-1\right),
\label{eq:rho}
\end{equation}
and
\begin{equation}
P=\frac{m_F^3}{12\pi}\left(\frac{\mu^3}{m_F^3}-
3\frac{\mu}{m_F}+2\right),
\label{eq:P}
\end{equation}
respectively.
Here $\mu$ stands for a chemical potential for the fermion field.

The number density of fermions is given by
\begin{equation}
n_F=\frac{m_F^2}{4\pi}\left(\frac{\mu^2}{m_F^2}-1\right).
\label{eq:nF}
\end{equation}


We assume the three-dimensional metric 
for a static circularly symmetric spacetime as
\begin{equation}
ds^2=-\alpha^2(r)dt^2+\beta^2(r)dr^2+
\frac{\gamma^2(r)}{\gamma_0^2}d\theta^2,
\label{eq:met}
\end{equation}
where $\alpha$, $\beta$, and $\gamma$ are functions of the radial
coordinate $r$ only. 
The constant $\gamma_{0}$ will be determined later 
(see Eqs.~(\ref{eq:EQ19}-\ref{eq:EQ20}) and below.). 
Because of the
coordinate invariance with respect to $r$, we have  a freedom in 
choosing the function $\beta(r)$. We will use this residual 
``gauge choice'' later.

We also assume that the complex scalar has 
a phase which is linear in the temporal coordinate:
\begin{equation}
\varphi=e^{-i\omega t}\varphi(r),
\end{equation}
where $\varphi(r)$ is a function of the radial coordinate $r$ only 
and $\omega$ is a constant.

For a static circularly symmetric spacetime,
$\rho$ and $P$ have only dependence on $r$.
Thus $\mu$ is also assumed to be a function of the radial coordinate
$r$ only.


Then the Einstein equation~(\ref{eq:EE}) implies the following
equations:
\begin{eqnarray}
\frac{1}{\beta\gamma}\frac{d}{dr}
\left(\frac{1}{\beta}\frac{d\gamma}{dr}\right)&=&
8\pi G\left[
-\frac{1}{\beta^2}\left(\frac{d\varphi}{dr}\right)^2-
\frac{\omega^2}{\alpha^2}\varphi^2-m_B^2\varphi^2-
\frac{\lambda}{2}\varphi^4-\rho\right]+C, \label{eq:EQb} \\
\frac{1}{\alpha\beta^2\gamma}\frac{d\alpha}{dr}
\frac{d\gamma}{dr}&=&
8\pi G\left[
+\frac{1}{\beta^2}\left(\frac{d\varphi}{dr}\right)^2+
\frac{\omega^2}{\alpha^2}\varphi^2-m_B^2\varphi^2-
\frac{\lambda}{2}\varphi^4+P\right]+C, \\
\frac{1}{\alpha\beta}\frac{d}{dr}
\left(\frac{1}{\beta}\frac{d\alpha}{dr}\right)&=&
8\pi G\left[
-\frac{1}{\beta^2}\left(\frac{d\varphi}{dr}\right)^2+
\frac{\omega^2}{\alpha^2}\varphi^2-m_B^2\varphi^2-
\frac{\lambda}{2}\varphi^4+P\right]+C. \label{eq:EQe}
\end{eqnarray}

The equation of motion for the scalar field~(\ref{eq:KG}) becomes
\begin{eqnarray}
\frac{\omega^2}{\alpha^2}\varphi&+&\frac{1}{\alpha\beta\gamma}
\frac{d}{dr}
\left(\frac{\alpha\gamma}{\beta}\frac{d\varphi}{dr}\right)-
m_B^2\varphi-\lambda\varphi^3=0. \label{eq:B1}
\end{eqnarray}

The equilibrium condition is reduced to
\begin{equation}
\frac{dP}{dr}=-\frac{1}{\alpha}\frac{d\alpha}{dr}\left(P+\rho\right).
\label{eq:F1}
\end{equation}


To simplify the equations, we make use of rescaled 
variables as
\begin{eqnarray}
\tilde{r}&=&m_Br,~~~\tilde{\varphi}=\sqrt{8\pi G}\varphi,~~~
\Lambda=\frac{\lambda}{8\pi Gm_B^2},~~~
\tilde{\omega}=\omega/m_B, \nonumber \\ 
\tilde{C}&=&C/m_B^2,~~~\tilde{\rho}=(8\pi G/m_B^2)\rho,~~~and~~~
\tilde{P}=(8\pi G/m_B^2)P.
\end{eqnarray}

We also rescale $\mu$ as
\begin{equation}
\tilde{\mu}=\mu/m_F.
\end{equation}


In order to study the limit of large self-inteaction, we rescale the
variables once more.
New variables are:
\begin{eqnarray}
r_{*}&=&\tilde{r}/\sqrt{\Lambda},~~~
\varphi_{*}=\sqrt{\Lambda}\tilde{\varphi},~~~
C_{*}=\Lambda\tilde{C}, \nonumber \\
\rho_{*}&=&\Lambda\tilde{\rho},~~~and~~~P_{*}=\Lambda\tilde{P}.
\end{eqnarray}

In the limit of large self-coupling, $\Lambda\rightarrow\infty$,
The equations~(\ref{eq:EQb}-\ref{eq:F1}) will be reduced to
\begin{eqnarray}
\frac{1}{\beta\gamma}\left(\frac{\gamma'}{\beta}\right)'&=&
-\frac{\tilde{\omega}^2}{\alpha^2}\varphi_{*}^2-\varphi_{*}^2-
\frac{1}{2}\varphi_{*}^4-\rho_{*}+C_{*}, \label{eq:EQ1}\\
\frac{1}{\beta^2}\frac{\alpha'}{\alpha}\frac{\gamma'}{\gamma}&=&
+\frac{\tilde{\omega}^2}{\alpha^2}\varphi_{*}^2-\varphi_{*}^2-
\frac{1}{2}\varphi_{*}^4+P_{*}+C_{*}, \label{eq:EQ2}\\
\frac{1}{\alpha\beta}\left(\frac{\alpha'}{\beta}\right)'&=&
+\frac{\tilde{\omega}^2}{\alpha^2}\varphi_{*}^2-\varphi_{*}^2-
\frac{1}{2}\varphi_{*}^4+P_{*}+C_{*}, \label{eq:EQ3} \\
\varphi_{*}&\times&\left(\frac{\tilde{\omega}^2}{\alpha^2}-
1-\varphi_{*}^2\right)=0, \label{eq:EQ4} \\
\left(\tilde{\mu}^2-1\right)&\times&\left(\tilde{\mu}'+
\frac{\alpha'}{\alpha}\tilde{\mu}\right)=0, \label{eq:EQ5}
\end{eqnarray}
where ${}'$ denotes the derivative with respect to $r_{*}$.
Here we have discarded the terms of order $1/\Lambda$ and higher.
Then the terms including ${\varphi_*}'$ disappear in the
equations.
This scheme to study the large self-coupling case was originally 
adopted by Colpi, Shapiro and Wasserman~\cite{CSW}.

\section{exact solutions}
\label{sec:3}

Now, let us solve the set of an algebraic equation and differential 
equations~(\ref{eq:EQ1}-\ref{eq:EQ5}).

Eq.~(\ref{eq:EQ4}) can be solved easily.
One solution for $\varphi_{*}$ is
\begin{equation}
\varphi_{*}^2=\frac{\tilde{\omega}^2}{\alpha^2}-1.
\label{eq:EQ6}
\end{equation}
We assume that this solution describes the configuration of the boson
field for $r_{*}<r_{*B}$.
Another solution for $\varphi_{*}$ is 
\begin{equation}
\varphi_{*}^2=0. \label{eq:EQ7}
\end{equation}
This solution is taken as the configuration of the boson
field for $r_{*}>r_{*B}$.
Note that $T^B{}^{\mu}_{\nu}=0$ for $r_{*}>r_{*B}$.
At the boundary $r_{*}=r_{*B}$, $\alpha$ takes the value
$\alpha=\tilde{\omega}$.


Eq.~(\ref{eq:EQ5}) can be solved easily, too.
One solution for $\tilde{\mu}^2$ is
\begin{equation}
\tilde{\mu}^2=\frac{\alpha_F^2}{\alpha^2},
\label{eq:EQ8}
\end{equation}
where $\alpha_F$ is a constant.
We assume that this solution describes the configuration of the 
fermion field for $r_{*}<r_{*F}$.
Another solution for $\tilde{\mu}^2$ is 
\begin{equation}
\tilde{\mu}^2=1. \label{eq:EQ9}
\end{equation}
This solution is taken as the configuration of the 
fermion field for $r_{*}>r_{*F}$.
Note that $T^F{}^{\mu}_{\nu}=0$,
or $\rho_{*}=P_{*}=0$ for $r_{*}>r_{*F}$.
At the boundary $r_{*}=r_{*F}$, $\alpha$ takes the value
$\alpha=\alpha_F$.


{}From Eqs.~(\ref{eq:EQ2}) and (\ref{eq:EQ3}), we find
\begin{equation}
\left(\frac{\alpha'}{\beta\gamma}\right)'=0.
\label{eq:EQ10}
\end{equation}
To solve this, we use the residual gauge choice on $\beta$.
For {\em simplicity}, we take
\begin{equation}
\alpha\beta\gamma=Kr_{*},
\label{eq:EQ11}
\end{equation}
where $K$ is a constant.
Actually, in order to obtain the solution, 
one can take an arbitrary choice.
This is because, as we will see later, we will solve the equations by
eliminating the radial coordinate variable. 
Then Eq.~(\ref{eq:EQ10}) is reduced to
\begin{equation}
\frac{d\alpha^2}{dr_{*}^2}=A,
\label{eq:EQ12}
\end{equation}
where $A$ is a constant.


{}From Eqs.~(\ref{eq:EQ2}) and (\ref{eq:EQ11}), 
we find%
\footnote{This equation~(\ref{eq:EQ13})
 is consistent with the other equations.}
\begin{equation}
\frac{1}{K^2}A\frac{\gamma\gamma'}{r_{*}}=
\frac{1}{2}\varphi_{*}^4+P_{*}+C_{*}.
\label{eq:EQ13}
\end{equation}
Here using Eq.~(\ref{eq:EQ12}), we obtain
\begin{equation}
\frac{A^2}{K^2}\frac{d\gamma^2}{d\alpha^2}=
\frac{1}{2}\varphi_{*}^4+P_{*}+C_{*}.
\label{eq:EQ14}
\end{equation}


Before exploring the explicit solution for the metric,
we rewrite the radial line element as
\begin{equation}
\beta dr=\sqrt{\frac{\Lambda}{m^2}}\beta dr_{*}=
\sqrt{\frac{\Lambda}{m^2}}\frac{Kr_{*}}{\alpha\gamma}
\frac{dr_{*}}{d\gamma}d\gamma=
\sqrt{\frac{\Lambda}{m^2}}
\frac{A}{K\alpha}\frac{1}{\frac{1}{2}\varphi_{*}^4+P_{*}+C_{*}}
d\gamma.
\label{eq:EQ15}
\end{equation}
Consequently, we get
\begin{equation}
\beta^2 dr^2=
\frac{\Lambda}{m^2}
\frac{A^2}{K^2\alpha^2}
\frac{1}{\left(\frac{1}{2}\varphi_{*}^4+P_{*}+C_{*}\right)^2}
d\gamma^2.
\label{eq:EQ16}
\end{equation}


First, we solve the exterior solution for $r_{*}>r_{*o}$,
where $r_{*o}=Max[r_{*B},r_{*F}]$.
Since the total energy momentum tensor vanishes
for $r_{*}>r_{*o}$, 
the solution of Eq.~(\ref{eq:EQ14}) is found to be
\begin{equation}
\alpha^2=\frac{A^2}{K^2C_{*}^2}\left(C_{*}\gamma^2-8GM_o\right),
\label{eq:EQ17}
\end{equation}
where $M_o$ is an integration constant.

The full line element for the exterior solution is obtained, by
substituting  Eqs.~(\ref{eq:EQ16}) and (\ref{eq:EQ17}) into
(\ref{eq:met}), as
\begin{equation}
ds^2=-\frac{A^2}{K^2C_{*}^2}
\left(CR^2-8GM_o\right)dt^2+
\frac{1}{CR^2-8GM_o}
dR^2+\frac{m^2}{\Lambda\gamma_0^2}R^2d\theta^2,
\label{eq:EQ19}
\end{equation}
where a new radial coordinate $R$ is defined as
\begin{equation}
R=\sqrt{\frac{\Lambda}{m^2}}\gamma.
\label{eq:EQ20}
\end{equation}
We should remember here that $C_{*}=C\Lambda/m^2$.
We take $\gamma_0=\sqrt{\frac{m^2}{\Lambda}}$ here to avoid an
appearance of a conical singularity~\cite{KS}. 
This choice makes $R$ a usual radial coordinate when
$\theta$ varies in the range $0\le\theta<2\pi$.

After rescaling $\frac{A}{KC_{*}}dt\rightarrow dt$,
we find the metric is precisely the same as
the well-known BTZ vacuum solution~\cite{BTZ}.
Thus we identify the constant $M_o$ with the BTZ mass 
{\em of the star}.


Next we must turn to solving the interior solution of the star.
We first consider the innermost region, $r_{*}<r_{*i}$,
where $r_{*i}=Min[r_{*B},r_{*F}]$.
{}From Eqs.~(\ref{eq:EQ6}), (\ref{eq:EQ8}) and (\ref{eq:EQ14}),
we obtain a differential equation:
\begin{equation}
\frac{A^2}{K^2}\frac{d\gamma^2}{d\alpha^2}=
\frac{1}{2}
\left(\frac{\tilde{\omega}^2}{\alpha^2}-1\right)^2+
F_{*}\left(\frac{\alpha_F^3}{\alpha^3}-
3\frac{\alpha_F}{\alpha}+2\right)+C_{*},
\label{eq:EQ21}
\end{equation}
where $F_{*}=(2G\Lambda m_F^3)/(3m_B^2)$.

The solution of this equation is given by
\begin{equation}
\frac{A^2}{K^2}\left(\gamma^2-\frac{B}{C_{*}}\right)=-\frac{1}{2}
\frac{\tilde{\omega}^4}{\alpha^2}-
\tilde{\omega}^2\ln\frac{\alpha^2}{\tilde{\omega}^2}+
\frac{1}{2}\alpha^2+
F_{*}\left(-2\frac{\alpha_F^3}{\alpha}-
6\alpha_F\alpha+2\alpha^2+6\alpha_F^2\right)+
C_{*}\alpha^2,
\label{eq:EQ22}
\end{equation}
where $B$ is a constant.


For the intermediate region $r_{*i}<r<r_{*o}$, we must consider two 
cases separately: for a) $r_{*B}>r_{*F}$ and b) $r_{*F}>r_{*B}$.


\noindent
{\em Case {\em a):} $r_{*B}>r_{*F}$}

In the region $r_{*F}<r_{*}<r_{*B}$,
one can solve the equation~(\ref{eq:EQ14}) with
(\ref{eq:EQ6}) and (\ref{eq:EQ9}) and get
\begin{equation}
\frac{A^2}{K^2}\left(\gamma^2-\frac{D_a}{C_{*}}\right)=-\frac{1}{2}
\frac{\tilde{\omega}^4}{\alpha^2}-
\tilde{\omega}^2\ln\frac{\alpha^2}{\tilde{\omega}^2}+
\frac{1}{2}\alpha^2+C_{*}\alpha^2,
\label{eq:EQ23}
\end{equation}
where $D_a$ is an integration constant.

At $r_{*}=r_{*F}$, when $\alpha=\alpha_F$,
the solutions~(\ref{eq:EQ22}) and (\ref{eq:EQ23}) must be
connected, thus we find
\begin{equation}
D_a=B.
\label{eq:EQ24}
\end{equation}

The value of $B$ is determined from the condition at
the outermost boundary of the star, $r_{*}=r_{*B}$,
where $\alpha=\tilde{\omega}$.
{}From the condition for a smooth connection of
the solutions~(\ref{eq:EQ17}) and (\ref{eq:EQ23}),
one can find 
\begin{equation}
D_a=B=8GM_o.
\label{eq:EQ25}
\end{equation}


\noindent
{\em Case {\em b):} $r_{*B}>r_{*F}$}

In the region $r_{*F}<r_{*}<r_{*B}$,
one can solve the equation~(\ref{eq:EQ14}) with
(\ref{eq:EQ7}) and (\ref{eq:EQ8}) and get
\begin{equation}
\frac{A^2}{K^2}\left(\gamma^2-\frac{D_b}{C_{*}}\right)=
F_{*}\left(-2\frac{\alpha_F^3}{\alpha}-
6\alpha_F\alpha+2\alpha^2+6\alpha_F^2\right)+
C_{*}\alpha^2,
\label{eq:EQ26}
\end{equation}
where $D_b$ is an integration constant.

At $r_{*}=r_{*B}$, where $\alpha=\tilde{\omega}$,
the solutions~(\ref{eq:EQ22}) and (\ref{eq:EQ26}) must be
connected, thus we find
\begin{equation}
D_b=B.
\label{eq:EQ27}
\end{equation}

The value of $B$ is determined from the condition at
the outermost boundary of the star, $r_{*}=r_{*F}$,
where $\alpha=\alpha_F$.
{}From the condition for a smooth connection of
the solutions~(\ref{eq:EQ17}) and (\ref{eq:EQ26}),
one can find 
\begin{equation}
D_b=B=8GM_o.
\label{eq:EQ28}
\end{equation}


In both cases ($a$ and $b$), we have found that the innermost solution
is given by
\begin{equation}
\frac{A^2}{K^2}\left(\gamma^2-\frac{8GM_o}{C_{*}}\right)=-\frac{1}{2}
\frac{\tilde{\omega}^4}{\alpha^2}-
\tilde{\omega}^2\ln\frac{\alpha^2}{\tilde{\omega}^2}+
\frac{1}{2}\alpha^2+
F_{*}\left(-2\frac{\alpha_F^3}{\alpha}-
6\alpha_F\alpha+2\alpha^2+6\alpha_F^2\right)+
C_{*}\alpha^2.
\label{eq:EQ29}
\end{equation}
In the region $r_{*i}<r_{*}<r_{*o}$,
the solutions are given by (\ref{eq:EQ23}) and (\ref{eq:EQ26})
with $D_a=D_b=8GM_o$ for {\em Case} a) and b), respectively.


In our solutions, the origin is located at $R=0$, or $\gamma=0$.
At the center of the star, we choose the following condition:
\begin{equation}
\frac{K^2\alpha_{0}^2}{A^2}
\left(\frac{1}{2}\varphi_{*0}^4+P_{*0}+C_{*}\right)^2=1,
\label{eq:EQ30}
\end{equation}
where $\alpha_{0}=\alpha(\gamma=0)$,
$\varphi_{*0}^2=\varphi_{*}^2(\gamma=0)=
\tilde{\omega}^2/\alpha_{0}^2-1$, and 
$P_{*0}=P_{*}(\gamma=0)=
F_{*}(\alpha_F^3/\alpha_0^3-3\alpha_F/\alpha_0+2)$. 

With this choice, one can find that the limit of ``no matter''
fields yields the three dimensional anti-de Sitter space with
no conical singularity~\cite{KS}.


Consequently, we obtain the relation between the BTZ mass of the
boson-fermion star and $\varphi_{*0}$ and
$\tilde{\mu}_0=\alpha_F/\alpha_0$ from Eq.~(\ref{eq:EQ29}):
\begin{equation}
\frac{8GM_o}{C_{*}}=\frac{\frac{1}{2}\varphi_{*0}^4+\varphi_{*0}^2-
\left(\varphi_{*0}^2+1\right)\ln\left(\varphi_{*0}^2+1\right)+
2F_{*}\left(\tilde{\mu}_0-1\right)^3-
C_{*}}{\left(\frac{1}{2}\varphi_{*0}^4+F_{*}
\left(\tilde{\mu}_0^3-3\tilde{\mu}_0+2\right)+C_{*}\right)^2}.
\label{eq:EQ31}
\end{equation}


The absence of boson and fermion fields is achieved in the limit
$\varphi_{*0}=0$ and $\tilde{\mu}_0=1$.
If one define the intrinsic mass $M_i$ as~\cite{KS}
\begin{equation}
M_i=M_o+\frac{1}{8G},
\label{eq:EQ32}
\end{equation}
the intrinsic mass $M_i$ vanishes when in the limit $\varphi_{*0}=0$ 
and $\tilde{\mu}_0=1$.

\section{particle numbers}
\label{sec:4}

The particle number $N_B$ of the boson field is given by
\begin{equation}
N_B=\int d^2x \sqrt{-g}\left|g^{tt}\right|i\left(
\varphi\partial_t\varphi^{*}-\varphi^{*}\partial_t\varphi\right).
\label{eq:EQ33}
\end{equation}
Therefore we can calculate the particle number of the boson
by making use of the above solution. It is given by
\begin{eqnarray}
N_B&=&
2\pi\int\frac{2m_B\tilde{\omega}}{8\pi G\Lambda}\varphi_{*}^2
\frac{1}{\alpha^2}\frac{\alpha\beta\gamma}{\gamma_0}dr \nonumber \\
&=&
\frac{m_B\tilde{\omega}}{2G\Lambda\gamma_0}
\int\left(\frac{\tilde{\omega}^2}{\alpha^2}-1\right)
\frac{1}{\alpha^2}Kr_{*}\sqrt{\frac{\Lambda}{m_B^2}}dr_{*} 
\nonumber \\
&=&
\frac{m_B\tilde{\omega}}{4G\Lambda\gamma_0}\sqrt{\frac{\Lambda}{m_B^2}}
\int_{\alpha_0^2}^{\tilde{\omega}^2}
\left(\frac{\tilde{\omega}^2}{\alpha^2}-1\right)
\frac{1}{\alpha^2}\frac{K}{A}d\alpha^2 \nonumber \\
&=&
\frac{m_B}{4G\Lambda\gamma_0}\sqrt{\frac{\Lambda}{m_B^2}}
\frac{K\alpha_0}{A}
\frac{\tilde{\omega}}{\alpha_0}
\left(\frac{\tilde{\omega}^2}{\alpha_0^2}-1-
\ln\frac{\tilde{\omega}^2}{\alpha_0^2}\right) \nonumber \\
&=&
\frac{1}{4Gm_B}\frac{\sqrt{\varphi_{*0}^2+1}
\left[\varphi_{*0}^2-
\ln\left(\varphi_{*0}^2+1\right)\right]}%
{\frac{1}{2}\varphi_{*0}^4+P_{*0}+C_{*}}.
\label{eq:EQ34}
\end{eqnarray}


The particle number of fermions is given by
\begin{eqnarray}
N_F&=&
\int d^2x \sqrt{-g}\frac{n_F}{\sqrt{-g_{tt}}} \nonumber \\
&=&
2\pi\int\frac{m_F^2}{4\pi}
\left(\frac{\alpha_F^2}{\alpha^2}-1\right)
\frac{\beta\gamma}{\gamma_0}dr \nonumber \\
&=&
\frac{m_F^2}{2\gamma_0}
\int\left(\frac{\alpha_F^2}{\alpha^2}-1\right)
\frac{1}{\alpha}Kr_{*}\sqrt{\frac{\Lambda}{m_B^2}}dr_{*} \nonumber \\
&=&
\frac{m_F^2}{2\gamma_0}\sqrt{\frac{\Lambda}{m_B^2}}
\int_{\alpha_0}^{\alpha_F}
\left(\frac{\alpha_F^2}{\alpha^2}-1\right)
\frac{K}{A}d\alpha \nonumber \\
&=&
\frac{m_F^2}{2\gamma_0}\sqrt{\frac{\Lambda}{m_B^2}}
\frac{K\alpha_0}{A}
\left(\frac{\alpha_F}{\alpha_0}-1\right)^2 \nonumber \\
&=&
\frac{m_F^2\Lambda}{2m_B^2}
\frac{\left(\tilde{\mu}_0-1\right)^2}%
{\frac{1}{2}\varphi_{*0}^4+P_{*0}+C_{*}} \nonumber \\
&=&
\frac{1}{4Gm_F}
\frac{3F_{*}\left(\tilde{\mu}_0-1\right)^2}%
{\frac{1}{2}\varphi_{*0}^4+P_{*0}+C_{*}}.
\label{eq:EQ35}
\end{eqnarray}


One can find that for small $\varphi_{*0}^2$ and $\tilde{\mu}_0-1$, 
the mass and the partcle
numbers are given approximately by
\begin{equation}
M_i\approx
\frac{\varphi_{*0}^4}{8GC_{*}}+
\frac{3F_{*}}{4GC_{*}}\left(\tilde{\mu}_0-1\right)^2\approx
m_BN_B+m_FN_F.
\label{eq:EQ36}
\end{equation}

\section{fermion stars}
\label{sec:5}

For a while, we will consider the fermion star,
which can be obtained by setting $\varphi_{*0}=0$.
In this case,
both maxima of the mass and fermion number as
functions of $\varphi_{*0}$
are located at $\tilde{\mu}_{0}=\tilde{\mu}_{0m}$.
$\tilde{\mu}_{0m}$ is given as
\begin{equation}
\tilde{\mu}_{0m}-1=\left(\frac{2C_{*}}{F_{*}}\right)^{1/3}
=\left(\frac{3C}{Gm_F^3}\right)^{1/3}.
\label{eq:EQ37}
\end{equation}

The maximum value of $M_{i}$, which we will denote as $M_{im}$
is found to be
\begin{equation}
M_{im}=\frac{1}{24G}
\frac{1}{\left[1+2\left(\frac{Gm_F^3}{3C}\right)^{1/3}\right]^2}+
\frac{1}{8G}.
\label{eq:EQ38}
\end{equation}

The maximum value of $N_{F}$, which we will denote as $N_{Fm}$
is found to be
\begin{equation}
N_{Fm}=\frac{1}{2Gm_F}
\frac{\left(\frac{Gm_F^3}{3C}\right)^{1/3}}%
{1+2\left(\frac{Gm_F^3}{3C}\right)^{1/3}}.
\label{eq:EQ39}
\end{equation}

The binding energy $E_{bm}=M_{im}-m_FN_{Fm}$ of
the fermion star with maximum fermion number is therefore
given by
\begin{equation}
E_{bm}=-\frac{1}{6G}
\frac{3\left(\frac{Gm_F^3}{3C}\right)^{2/3}-1}%
{\left[1+2\left(\frac{Gm_F^3}{3C}\right)^{1/3}\right]^2}.
\label{eq:EQ40}
\end{equation}

If the value of the cosmological constant is larger than $C_{crit}$,
this state is energetically unfavorable 
because $E_{bm}$ is positive.
The critical value is
\begin{equation}
C_{crit}=\sqrt{3}Gm_F^3.
\label{eq:EQ41}
\end{equation}

It is worth noting that the value of maximum BTZ mass 
$(M_{om}=M_{im}-\frac{1}{8G})$ is positive 
for all finite values of $C$. In particular, in the limit
$C=0$, the value of $M_{om}$ vanishes. The behavior of 
$M_{om}$ for small $C$ is given by $M_{om}\propto
C^{2/3}$.
The value of $N_{Fm}$ in the limit $C=0$ is $1/(4Gm_F)$.

If the minimum binding energy is positive,
the fermion star is no longer energetically favorable.
Thus when $C>\sqrt{3}Gm_F^3$, there is no static configuration of
the fermion star.

\section{stablity of boson-fermion stars}
\label{sec:6}

The stability of four-dimensional boson-fermion stars~\cite{BFS}
has been studied by several authors.
The perturbative stability can be judged by
the line of zero eigenvalue about particle numbers in the
($x$, $y$) plane, where we denote $\varphi_{*0}$ and $\tilde{\mu}_0-1$
as $x$ and $y$, respectively.
The condition of zero eigenvalue is
\begin{equation}
\left|
\begin{array}{cc}
\frac{\partial N_B}{\partial x} &
\frac{\partial N_B}{\partial y} \\
\frac{\partial N_F}{\partial x} &
\frac{\partial N_F}{\partial y}
\end{array}\right|=0.
\label{eq:EQ42}
\end{equation}

In our solutions, the condition can be reduced to
\begin{equation}
x^2\left(-6C_{*}+4x^2+x^4+3F_{*}y^3\right)
+\left(2C_{*}-4x^2-3x^4-F_{*}y^3\right)\ln\left(1 +x^2\right)=0,
\label{eq:zero}
\end{equation}
where $x\equiv \varphi_{*0}$ and $y\equiv \tilde{\mu}_0-1$.

\begin{figure}
\centering
\epsfbox{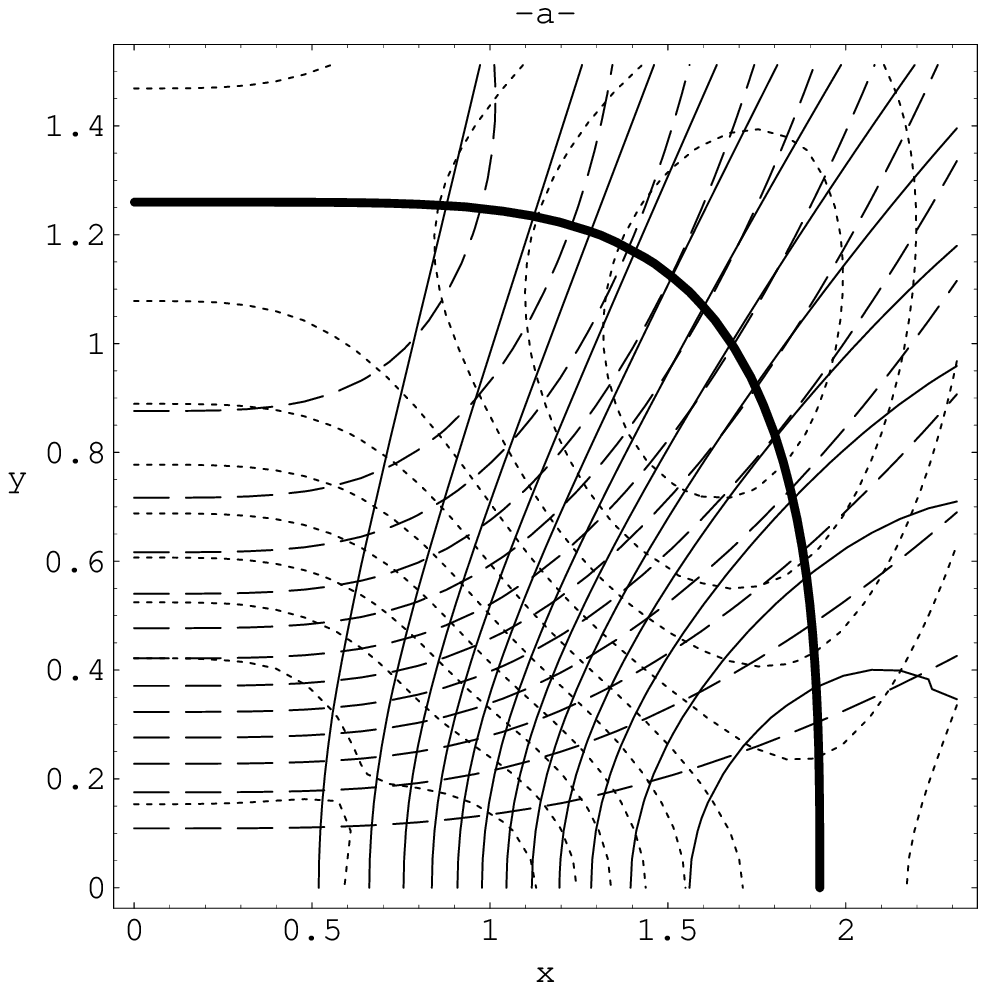}

\epsfbox{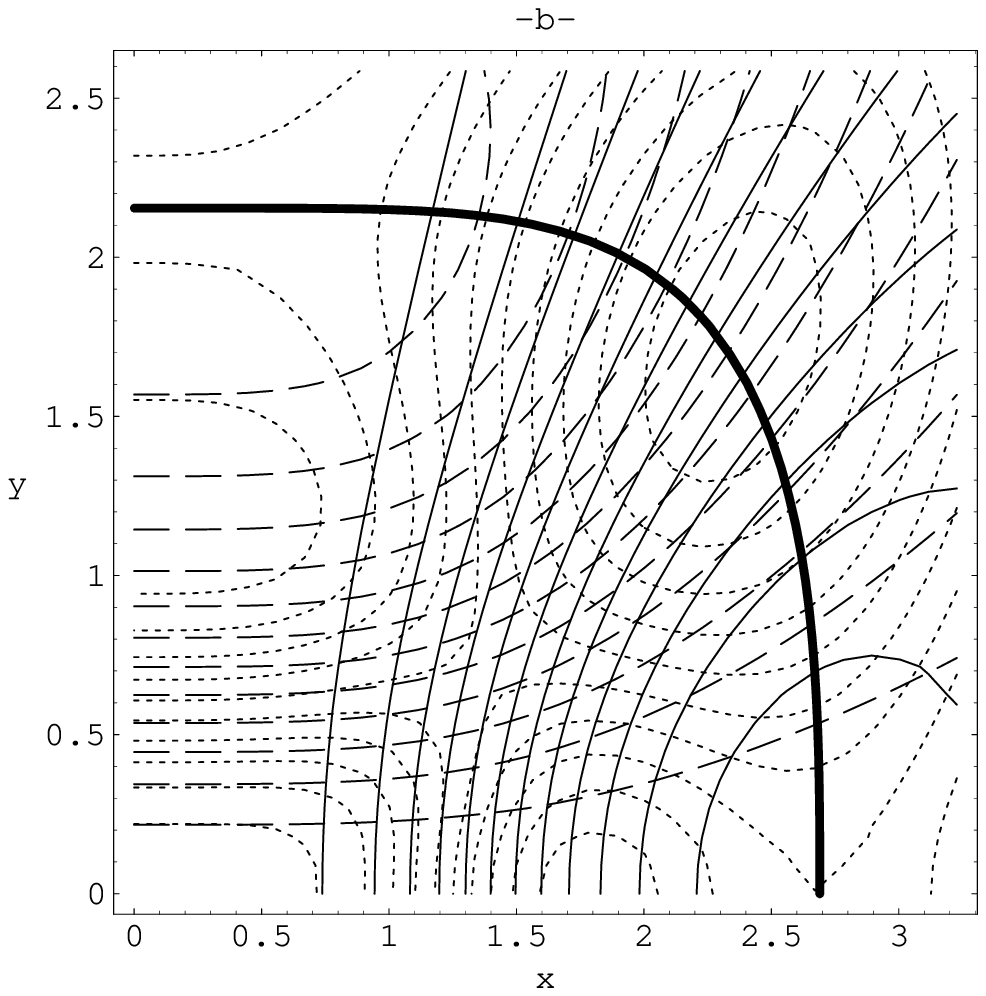}

\epsfbox{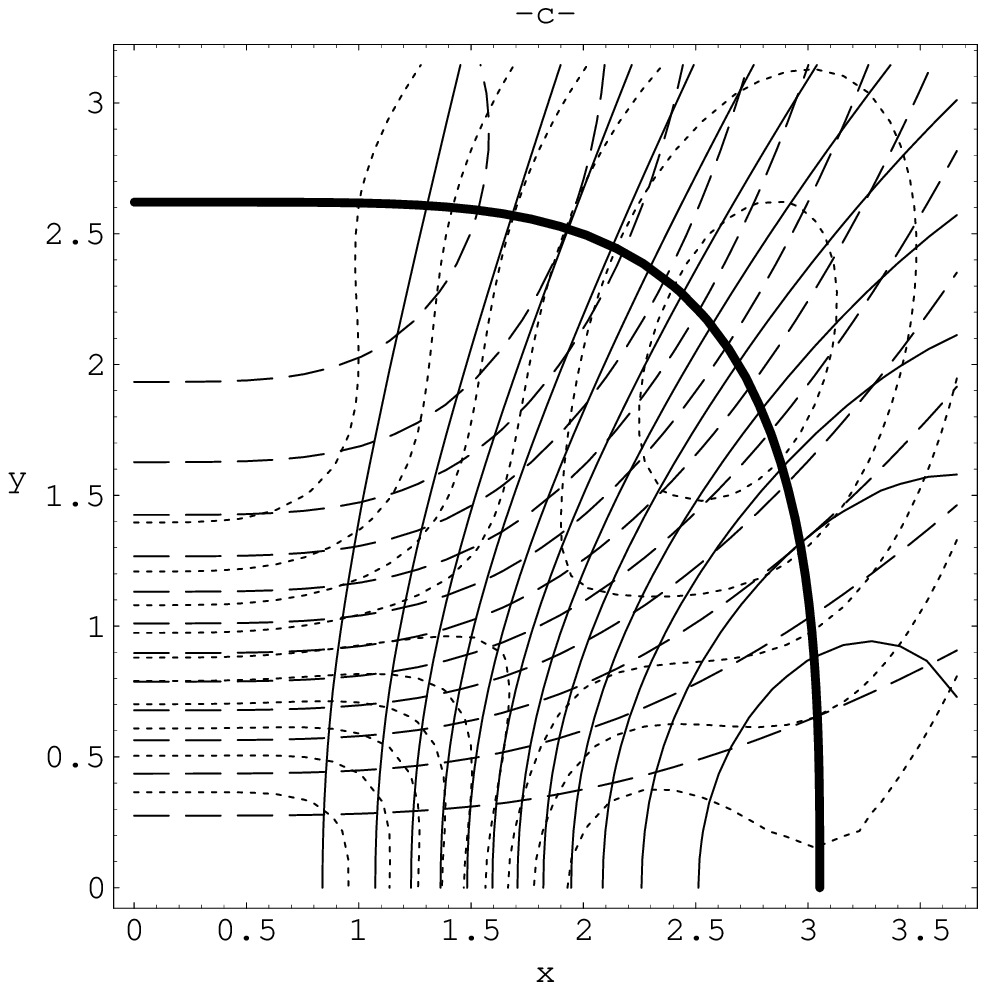}

\caption{Contour lines of constant $N_B$ (solid lines)
and $N_F$ (broken lines) in the 
($x$, $y$) plane, where $x=\varphi_{*0}$ and $y=\tilde{\mu}_0-1$. 
The bold line shows the line of zero eigenvalue (see text). 
Contour lines of the binding energy is also shown by dotted lines. In
these figures, $F_{*}$ is set to unity, 
and $C_{*}$ is ($a$) $1$, ($b$) $5$, and ($c$) $9$.}
\label{fig1}
\end{figure}

In FIG.~\ref{fig1} we show the line of zero eigenvalue and contour 
lines of constant $N_B$ and $N_F$ in the  ($x$, $y$) plane.
The perturbatively stable solutions for boson-fermion stars
lie on the left of the line of zero eigenvalue.
The solutions on the line is absolutely stable,
in the sense that the stars does not exhibit radial oscillations.


In our three-dimensional model,
the binding energy of the star can take positive value 
even in the left region of the line of zero eigenvalue.
Thus the minimum of the binding energy is not always a global minimum.

The contour lines of the binding energy $E_b=M_i-m_BN_B-m_FN_F$
is also plotted in  FIG.~\ref{fig1}.

The minimum binding energy point is given by the crossing point
of zero-eigenvalue line (\ref{eq:zero}) and the following line:
\begin{eqnarray}
& &C_{*} y \left[-2C_{*}+2x^2+x^4+
F_{*} y \left(6+3y+y^2\right)\right]\nonumber \\
&=&\left[\frac{1}{2}x^4+F_{*} y^2
\left(3+y\right)+C_{*}\right]\left[4x^2+3x^4+
F_{*}y^3-2\left(y+2\right)x^2\sqrt{1+x^2}\right],
\end{eqnarray}
where $x\equiv \varphi_{*0}$ and $y\equiv \tilde{\mu}_0-1$.

Unfortunately, the values of $x=\varphi_{*0}$ and 
$y=\tilde{\mu}_{0}-1$ which yield the configuration of the minimum 
binding energy cannot be obtained analytically.

\begin{figure}
\centering
\epsfbox{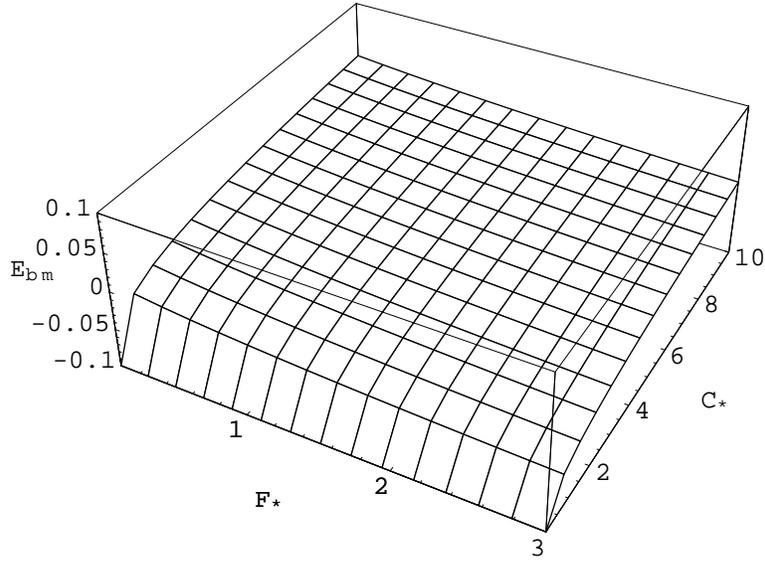}
\caption{The value of minimum binding energy of a boson-fermion star
as a function of $F_{*}$ and $C_{*}$.}
\label{fig2}
\end{figure}

The value of minimum binding energy $E_{bm}$ is plotted as a 
function of $F_{*}$ and $C_{*}$ in FIG.~\ref{fig2}.
One can see that the minimum binding energy becomes positive
for sufficiently large $C_{*}$ and small $F_{*}$.

\begin{figure}
\centering
\epsfbox{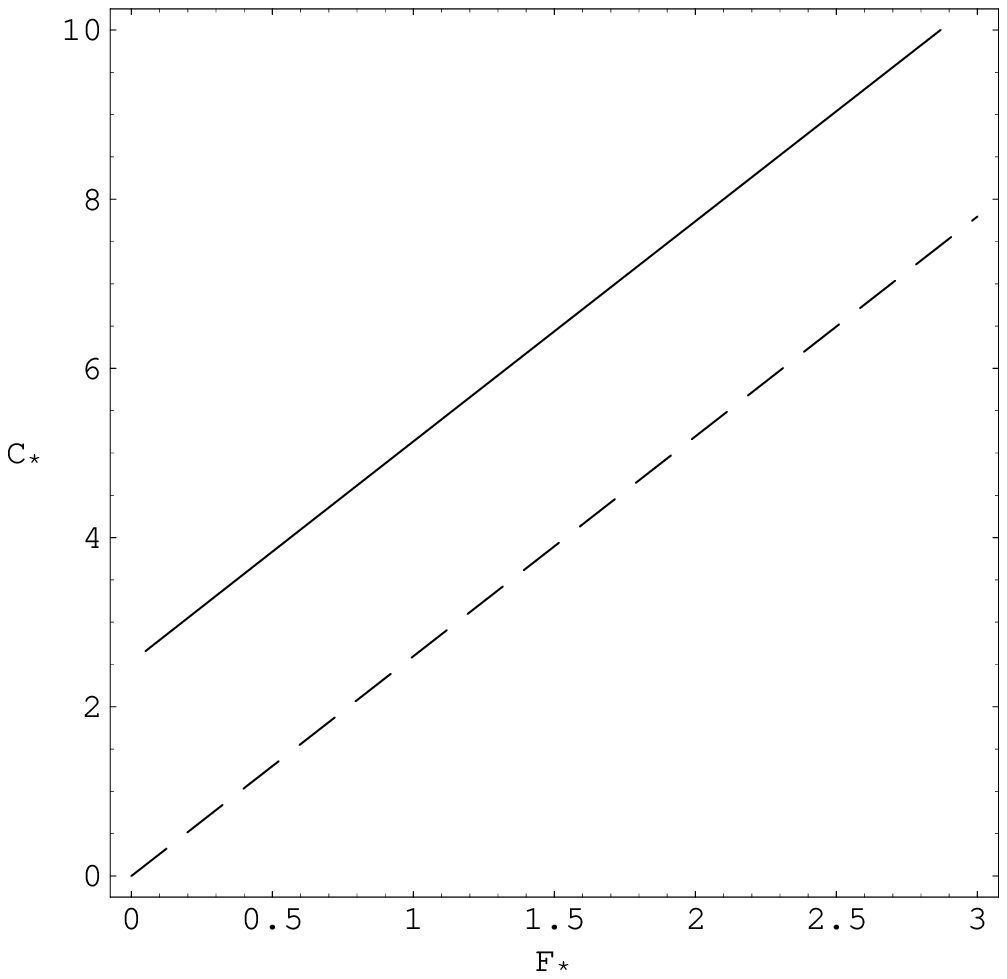}
\caption{The line of zero minimum binding energy
in the ($F_{*}$, $C_{*}$) plane (the solid line).
The broken line shows the line of
$C_{*}=\frac{3\sqrt{3}}{2}F_{*}$.}
\label{fig3}
\end{figure}

The critical line, i.e., the line of zero minimum binding energy
is shown in FIG.~\ref{fig3}.
In the limit of $F_{*}=0$, the line ends with 
$C_{*}=C_{*~crit}$, where $C_{*~crit}\approx 2.5268$ 
is the critical value obtained in~\cite{KS} for boson stars.
This fact is trivial because the limit $F_{*}=0$ means 
no contribution of fermions.

The broken line in FIG.~\ref{fig3} shows the line of
\begin{equation}
C_{*}=\frac{3\sqrt{3}}{2}F_{*}.
\end{equation}
It is obvious that the critical value~(\ref{eq:EQ41}) can be found
for fermion stars, when the contribution of fermions is dominant.

If the minimum binding energy is positive,
a boson-fermion star with any particle numbers has
positive binding energy.
Therefore, for arbitrary values of $F_{*}$,
no stable boson-fermion star exists for sufficiently large value
of $C_{*}$. The critical value is given approximately by
\begin{equation}
C_{*~crit}(F_{*})\approx 2.5268+\frac{3\sqrt{3}}{2}F_{*}.
\label{eq:EQ45}
\end{equation}

Finally, we give the mass and particle numbers in a special limit,
$C_*<<1$ and $F_{*}<<1$.
In this case, the values of $x=\varphi_{*0}$ and $y=\tilde{\mu}_0-1$
which yield the minimum binding energy can be approximately solved.
These values, we denote them as $x_m=\varphi_{*0m}$ and 
$y_m=\tilde{\mu}_{0m}-1$, is given by
\begin{equation}
x_m=\varphi_{*0m}\approx 
\left(24C_{*}\left(1-\frac{3}{2}F_{*}\right)\right)^{1/6},~~~~~
y_m=\tilde{\mu}_{0m}-1\approx 
\left(3C_{*}\right)^{1/3}.
\end{equation}

These values lead to the mass $M_{im}$, particle number of bosons
$N_{Bm}$, and particle number of fermions $N_{Fm}$
of the boson-fermion star of the minimum binding energy.
The approximate values of them are
\begin{eqnarray}
GM_{im}&\approx &\frac{1}{8}+
\frac{C_{*}^{2/3}}{32\sqrt[3]{3}\left(
\frac{3}{2}F_{*}+\left(1-\frac{3}{2}F_{*}\right)^{2/3}
\right)^{2}}\nonumber \\
Gm_BN_{Bm}&\approx &
\frac{\left(1-\frac{3}{2}F_{*}\right)^{2/3}}{4\left(
\frac{3}{2}F_{*}+\left(1-\frac{3}{2}F_{*}\right)^{2/3}
\right)}\nonumber \\
Gm_FN_{Fm}&\approx &
\frac{\frac{3}{2}F_{*}}{4\left(
\frac{3}{2}F_{*}+\left(1-\frac{3}{2}F_{*}\right)^{2/3}
\right)}.
\end{eqnarray}
One can see that $Gm_BN_{Bm}+Gm_FN_{Fm}=1/4$ in this case.
It is worth noting that the BTZ mass of the boson-fermion star
approaches zero in the limit of $C_{*}=0$.


\section{discussion}
\label{sec:7}

We have obtained exact solutions describing boson-fermion
stars with very large self-coupling constant in $(2+1)$ dimensions.
There is a critical value for $C_{*}=(\lambda/(8\pi Gm^4))C$,
where $C$ is the absolute value of the (negative) cosmological 
constant. 
For $C_{*}>C_{*~crit}(F_{*})$ the binding energy of
any solution is positive,
the boson-fermion star configuration is not energetically favorable.

To study the $(2+1)$ dimensional boson star, with a finite 
value of the self-coupling or direct couplings between bosons and
fermions, we must perform numerical analyses.
The present study of the exact solution is the ground for studying
the properties of the boson-fermion stars in general cases.

For a finite self-coupling case, the external metric cannot be
described by the exact BTZ metric. We expect that the external metric
will approach the black hole metric in the large distance 
asymptotically if the global attractive force due to the cosmological
constant is sufficiently effective.
On the other hand, another types of the external metric will also be
expected, which exhibit different behaviors from the BTZ metric.
The possibility is worth studying.

In the present paper, we have discussed only static configurations.
We are interested also in the process of formation of the 
boson-fermion stars. 
The formation of stars or black holes (or possible naked 
singularities) will be clarified by the investigation of 
time-dependent processes. 
The quantum mechanical processes,
including even the tunneling process, may play an important role
in the creation of small boson-fermion stars.
This subject will be considered in a separate research in the future.

The time-dependent solution describing radial pulsations in 
$(2+1)$ dimensions can be studied first of all.
It will be discussed in elsewhere.

\section*{Acknowledgement}
The authors would like to thank the referee for comments.



\newpage

\newpage

\begin{references}
\bibitem{FTD} K.~Shiraishi, 
Prog.\ Theor.\ Phys.\ {\bf 77}, 1253 (1987).
The present results are obtained from the
thermodynamic potential~(4.9)
only with the $\ell=0$ term, $d_0=1$ and $\omega_0=m_F$, 
and divided by $2$.
(This factor $2$ has been missed throughout that paper.)
\end{references}
\end{document}